\documentclass{mn2e}
\usepackage{graphicx}
\begin{document}
\title{Gravitational Wave detection through microlensing?}
\author[Ragazzoni, Valente \& Marchetti]{Roberto Ragazzoni$^{1,2}$, Gianpaolo Valente$^{3,4}$, Enrico Marchetti$^5$\\
$^1$INAF, Astrophysical Observatory of Arcetri, largo
Enrico Fermi 5, I--50125 Firenze (Italy) {\tt
ragazzoni@arcetri.astro.it} \\
$^2$Max Plank Institut f\"ur
Astronomie, K\"onigstuhl 17, D--69117 Heidelberg (Germany) \\
$^3$Dipartimento di Fisica ``G. Galilei'', Universit\`{a} degli
Studi di
Padova, via Marzolo 8, I--35131 Padova (Italy) {\tt valente@pd.infn.it}\\
$^4$INFN, Sezione di Padova  \\
$^5$European Southern
Observatory, Karl--Schwarzschild--Str. 2, D-85748 Garching bei
M\"unchen (Germany) {\tt emarchet@eso.org} }

\maketitle

\begin{abstract}
It is shown that accurate photometric observations of a relatively
high--magnification microlensing event ($A\gg 1$), occurring close
to the line of sight of a gravitational wave (GW) source,
represented by a binary star, can allow the detection of subtle
gravitational effects. After reviewing the physical nature of such
effects, it is discussed to what extent these phenomena can
actually be due to GWs. Expressions for the amplitude of the
phenomena and the detection probability are supplied.
\end{abstract}


\section{Introduction}

Gravitational Waves (GWs) are predicted by General Relativity (GR) and their existence
has been indirectly proven by binary pulsar timing (Hulse \&
Taylor, 1975; Taylor, Fowler \& McCulloch, 1979). GWs, as received
on the Earth from any astrophysical source, produce extremely
small effects and no GW has been detected, yet.

The effect of GWs on some astrophysical measurable quantities have
been the subject of  studies and proposals: but, as a result, only
upper bounds on the strength of GWs have been posed.
{\it Scintillation} of the starlight by the GWs focussing of the
electromagnetic radiation toward the Earth (Labeyrie, 1993;
Bracco, 1997) have been proposed (Zipoy, 1966) to place an upper
limit on the theoretically predicted GW background, in analogy
with the Cosmic Microwave Background Radiation.
To detect the same stochastic background of the GWs,
the deflection of electromagnetic beams has been studied by
several authors (Linder, 1988; Bar--Kana, 1996; Pyne, Gwinn \&
Birkinshaw, 1996) giving, again, only upper bounds. The same has
been made in studies on the time--delay in lensed Quasar images
(Frieman, Harari \& Surpi, 1994).
It is worth noting a proposal to
find GWs from SuperNovae (Fakir, 1993, 1994b) and to discriminate
between different gravitational theories (Faraoni, 1996; Bracco \&
Teyssandier, 1998), observing light deflection from the GWs
themselves. Most of these techniques rely on the basic idea that
the strength of a GW, while extremely low on the Earth, would be
noticeably larger if detected closer to the source and
that the deflection angle of a light beam, interacting with such a
GW is of the order of the metric perturbation in the region of
closest approach. However, it has been recently pointed out that,
at least in the standard GR framework and up to a certain degree
of approximation, such a statement is incorrect (Bracco 1998,
Damour \& Esposito--Farese 1998, Kopeikin et al. 1999, for a brief discussion see also Crosta, Lattanzi \&
Spagna, 1999).

In our Galaxy the most noticeable (and predictable) sources of GWs
are
binary stars (Lipunov, Postnov \& Prokorov, 1987), in particular,
 W--UMa stars (Mironovskii, 1966).

A microlensing event occurs when, by chance, a precise alignment between a
background
source and a deflecting mass is experienced by the observer (Paczi\'nsky,
1986). The gravitational perturbation generated by a binary star
very close to the line
of sight of the event, slightly deflects light ray trajectories (Durrer, 1994),
introducing
a small distortion in the microlensing alignment. This effect translates into a
modulation of the light amplification that is synchronous with the binary star.

While we note that a similar idea has been
proposed to describe some light variation in Quasar microlensing (Schild \& Tompson, 1997;
Larson \& Schild, 2000), here we intend to focus our attention to the particular case of
the perturbation by a binary star of a microlensing event, inside our Galaxy.

The technique that will be described in the following sections could lead,
in our opinion, to an unambiguous detection of such subtle gravitational effects from a
stable source.
Even if such a detection has to occur just during  a microlensing alignment,
the GW source can be studied and observed well outside  the
microlensing time--span. It is remarkable
that the perturbation of the microlensing is not the most
sensitive--to--misalignment astronomical phenomenon
one can conceive.
In fact, microlensing, including light interference effects
(Deguchi \& Watson, 1986; Ulmer \& Goodman, 1995; Jaroszi\'nsky \& Paczi\'nsky,
1995) could lead to an even greater detection probability of such elusive phenomena.

The role of the approximations employed to get the solutions for
the light ray propagation equations is crucial, in order to
determine to what extent the deflection angle may be directly
ascribed to genuine GWs. Correspondingly, how the measurements of
such small deflection angles and their time behaviour coud lead to
new insight on gravitational theory and/or GW detection, remains
questionable.

Hereafter, we intend to review the contributions to the light deflection angle
deriving from some theoretical assumptions, based
on different degrees of approximations,
together with a discussion of their physical relevance.
It is worth noting that, in the calculations so far published, the role of the first radiative term,
if any, has still to be clarified.
Then, we compute the
effects of the light deflection on the microlensing alignment and
hence on the perturbation of the
observed light--curve; we carry out a numerical example with some
reasonable figures and, finally, we roughly estimate the probability to
detect such an event within the framework of the existing campaigns to
detect and follow--up microlensing events.

\section{Deflection of light due to a binary star close to the line of sight}

The section is organized as follows: after setting some preliminary definitions,
mostly of geometrical nature, we first give an estimate of the perturbation $h$ determined
from a GW source on the flat space-time Minkowskian metric. Then, we deal with light angle deflection
calculation under different approximations: a quasi static, post--Newtonian case, where the speed
of the bodies in the binary star are considered to be zero; a post--Minkowskian treatment, where
 body speeds are not negligible, using a second order light deflection formalism and, finally,
we report the results by Fakir (1994a).
  After expanding the deflection angle expressions up to any power of the impact parameter $d$
  (hence not relying just on the leading term of the series) we arrange them, with the aim to
   compare their relative strength and to discuss the approximation required to establish
    the nature of the radiative component of the perturbation.

\subsection{Preliminary definitions}

\begin{figure}
\includegraphics[width=84mm]{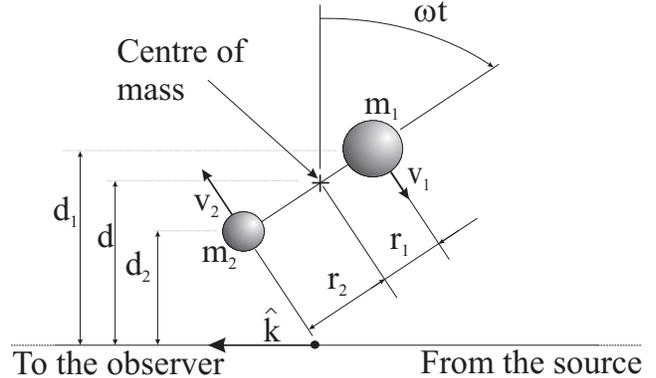}
  \caption{ \label{gwsl} An isolated binary star, characterized by
 masses $m_1$ and $m_2$, in circular orbits with radius $r_1$ and
 $r_2$ around their common centre of mass and angular speed
 $\omega$, is approached by a light beam with an impact parameter
 $d$. Versor $\hat k$ characterizes light propagation direction.
Other geometrical quantities are indicated in the figure. }
\end{figure}

Let us consider (see Fig.~\ref{gwsl}) a double star, whose
components have masses $m_1$ and $m_2$, in circular orbit around
the common center of mass. The two stars are separated by a
distance $\rho = r_1 +r_2$. Special care is to be given, here, to
the relativistic definition of
 centre of mass, $\vec{r}_{\mathrm{ cm}}$,
which  is well known in the non--interacting case; denoting by
${\cal E}_i$ the i--th body energy, the formula is the following:
\begin{equation}  \label{cm}
\vec{r}_{\mathrm{ cm}} = \frac{{\cal E}_1 \vec{r_1} + {\cal E}_2 \vec{r_2}}{{\cal E}_1  + {\cal E}_2 }\;.
\end{equation}
Formally, Eq.~(\ref{cm}) can be recovered from the classical
mechanics one,  just replacing all the masses with their
relativistic counterparts:
\begin{equation} \label{massarel}
m_i^* = \frac{m_{0i}}{\sqrt{1-v_i^2/c^2}}\;, \;\;\; i=1,2\,,
\end{equation}
where $m_{0i}$ denote the rest masses of the two component stars. For interacting bodies, however, in Eq.~(\ref{cm}),
a binding energy
\begin{equation}
 W_b= - \frac{G m_1 m_2}{2 \rho}\,,
\end{equation}
has to be taken into account. Imitating what we have just done in
Eq. (\ref{massarel}), one has to replace the mass terms in the
classical formula with the more complicated ones (see for instance
Landau \& Lifshitz, 1971, problem~2, section~106):
\begin{equation} \label{massarelint}
m_i = m_i^* + {\displaystyle \frac{W_b}{c^2}}\;, \;\;\;\; i=1,2\,.
\end{equation}
Setting the origin of our reference frame in the centre of mass, by definition:

\begin{equation}
m_1r_1 = m_2 r_2\,
\end{equation}
and one can define the time--dependent impact parameters of the two bodies as follows:
\begin{equation} \label{d1d2}
\left\{
\begin{array}{rcl}
d_1 & = & d + r_1 \cos\omega t \\
d_2 & = & d - r_2 \cos\omega t
\end{array}
\right. \;,
\end{equation}
where it is assumed that the eccentricities of the orbits are zero
({\em i.e.} circular orbit) and the body motion is characterized
by an angular speed $\omega$. One can also define the velocity
components of the two bodies, along the approaching light beam
direction, as $V_{ik} = \vec{v_i} \cdot \widehat{k}$, so that:
\begin{equation}
\left\{
\begin{array}{rcl}
V_{1k} & = & - \omega r_1 \cos \omega t \\
V_{2k} & = & \omega r_2 \cos \omega t
\end{array}
\right.\;,
\end{equation}
where $\widehat{k}$ is a versor aligned with the unperturbed light propagation toward the observer.
As usual, the system total mass is denoted by $M=m_1+m_2$ and its reduced mass by:
\begin{equation}
\mu = \frac{m_1m_2}{m_1+m_2}\,.
\end{equation}
The orbital period $P$ is related both to the binary angular speed $\omega$
and to the binary frequency $f$ by the usual relations:
\begin{equation}
\omega = \frac{2\pi}{P} = 2\pi f .
\label{eqomega}
\end{equation}
Recalling Eq.~(\ref{eqomega}), the third Kepler's law
\begin{equation}
\frac{\rho^3}{P^2 M} = \frac{G}{4 \pi^2}
\end{equation}
can be cast in the other useful form:
\begin{equation}
\omega^2 \rho^3 = G M\,.
\end{equation}

Let us assume that the binary star radiates power essentially in the form of GWs at the
single quadrupole frequency $f_G=2f$ (or with angular phase speed
$\omega_G = 2 \omega$) with an intensity given by

\begin{equation}
W= \frac{32G}{5c^5} \mu^2 \rho^4 \omega^6\,.
\label{wgrande}
\end{equation}
On the other hand, the energy density $w$ carried out by a GW is given by
\begin{equation}
w = \frac{c^3}{16\pi G} \left( \frac{\partial h}{\partial t} \right)^2\;,
\label{wpiccola}
\end{equation}
where $h$ is the dimensionless amplitude of the GW perturbation of the metric
(Derouelle \& Piran, 1983; Hawking \& Israel, 1979; Weinberg, 1972).
In the particular case of a monochromatic GW with phase angular speed
$\omega_G$ we have
\begin{equation}
h=h_0 \sin (\omega_G t)
\end{equation}
and the quadratic  term appearing on the right hand side of
Eq.~(\ref{wpiccola}) becomes
\begin{equation}
\left( \frac{\partial h}{\partial t} \right)^2 = \frac{h_0^2 \omega_G^2}{2}
\left[ 1 + \cos (2 \omega_G t) \right]\;.
\end{equation}
After averaging it for a time span much larger than $1/f_G$ one gets:
\begin{equation}
\left\langle \left( \frac{\partial h}{\partial t} \right)^2 \right\rangle = \frac{h_0^2 \omega^2_G}{2}
\end{equation}
and substituting it for the binary rotation angular speed figure
\begin{equation}
\left\langle \left( \frac{\partial h}{\partial t} \right)^2 \right\rangle = 2 h_0^2 \omega^2\,,
\end{equation}
so that Eq.~(\ref{wpiccola}) becomes:
\begin{equation}
w = \frac{c^3 h_0^2 \omega^2}{8\pi G}\,.
\label{wdensity}
\end{equation}

Assuming an isotropic energy distribution around the binary
star, the energy density at a distance $d$ from the GW source is:
\begin{equation}
w = \frac{W}{4\pi d^2}
\end{equation}
and,  using Eq.~(\ref{wgrande}) and Eq.~(\ref{wdensity})
one finally obtains:
\begin{equation} \label{equa20}
h_0 = \frac{8}{\sqrt{5}} \frac{G \omega^2 \mu \rho^2}{c^4 d}\,.
\end{equation}
From the 3rd Kepler's law
\begin{equation}
\rho^2 = \frac{G^{2/3} M^{2/3}}{\omega^{4/3}}\;,
\end{equation}
 hence
\begin{equation}
h_0 = \frac{8}{\sqrt{5}} \frac{G^{5/3} \omega^{2/3} \mu M^{2/3}}{c^4 d}\;.
\label{accazero}
\end{equation}
A numerical approximation of Eq.~(\ref{accazero}) can be given in MKS units:
\begin{equation}
h_0 \approx 4.85 \cdot 10^{-51} \frac{\omega^{2/3} \mu M^{2/3}}{d}
\end{equation}
and in astrophysical ones:
\begin{equation}
h_0 \approx 1.77 \cdot 10^{-14}
\frac{\left(\mu/M_\odot\right) \left(M/M_\odot
\right)^{2/3}}{\left(P/{\mathrm{ days}}\right)^{2/3} \left(d/{\mathrm{ AU}} \right)}\;.
\end{equation}
At one GW wavelength $\Lambda$ given by
\begin{equation} \label{equa25}
\Lambda = \frac{2\pi c}{\omega_G} = \frac{\pi c}{\omega}\,,
\end{equation}
the metric perturbation $h_0$  becomes:
\begin{equation}
h_0(\Lambda) = \frac{8}{\pi \sqrt{5}} \frac{G^{5/3} \omega^{5/3}
\mu M^{2/3}}{c^5}\,.
\end{equation}

\subsection{Calculation technique}

Hereafter, for the sake of simplicity, let us assume that the
deflection of a light--ray due to a GW source is confined to a
very small region, close to the point of minimum distance between
the ligth--ray and the GW source. As a consequence, such a
perturbation is  characterized by the deflection angle, since we
can assume that the interaction is essentially {\em concentrated}
at the minimum impact  point. In what follows let us denote such
an angle by $\alpha$ with various pedices or special signs, in
order to distinguish the various degrees of approximation used to
derive such a value.

We rewrite Eq.~68 of Kopeikin \& Sch\"afer (1999) for the bending angle
of the light ray $\tilde \alpha$, under the assumption that the impact parameter
$d$ is negligible with respect to the distance of the deflecting masses from the observer,
obtaining:

\begin{equation} \label{kopeikin68}
\tilde\alpha = -\frac{4G}{c^2} \sum_{i=1}^2 \frac{m_i\left(1-V_{ik} \right)}{d_i}\,.
\end{equation}
We stress that such a result is the same obtained by
Pyne \& Birkinshaw (1993) in their Eq.~45. Furthermore, let us rewrite the
total bending angle $\Delta \phi$ as the sum of several deflection
angle contributes:

\begin{equation}
\Delta \phi=\alpha_{\mathrm{PN}}+\alpha_{\mathrm{PM}} + \alpha_{\mathrm{PPN}} + \alpha_{\mathrm{F}} + \ldots
\end{equation}
where $\alpha_{\mathrm{PN}}$ refers to the Post--Newtonian formalism, that is to the
linearized Einstein field equation, in the slow motion approximation or, in other words,
approximating the light deflection angle
in Eq.~(\ref{kopeikin68}) for $V_{ik} \rightarrow 0 $ and, with a certain abuse of language:

\begin{equation}
\alpha_{\mathrm{PM}} =  \tilde\alpha - \alpha_{\mathrm{ PN}}\,,
\end{equation}
since we are mainly interested in the new effects arising from the
motion of the stars in the binary system under consideration.
Actually,  the term post--Minkowskian usually refers to the whole
amount given by $\tilde\alpha$; for a thorough discussion see
Thorne (1987).
Furthermore, a post--post--Newtonian term $\alpha_{\mathrm{ PPN}}$ is produced by second order
expansion term in $G$, while $\alpha_{\mathrm{ F}}$ refers
to the deflection effect claimed by Fakir (1994a) and criticized by Kopeikin et al. (1999)
and by Damour \& Esposito--Farese (1998).

In order to have an idea of the order of magnitude for the terms in Eq.(\ref{kopeikin68}), we just note that, by replacing $m$ and $d$ by the solar values, one obtains for the leading term the well known deflection angle at the edge of the Sun, namely $\approx 1.75"$.

The approach used here is to compute the deflection angle by writing the impact parameters $d_1$ and
$d_2$ of the two masses in the binary star, as a function of the average common
impact parameter $d$ (see Eq.~\ref{d1d2}). The result is then expanded in a series of $1/d$ terms.
The following relations will be used throughout this paper:
\begin{equation} \label{v27}
\left\{
\begin{array}{l}

{\displaystyle \frac{1}{1\pm \varepsilon}} \approx 1\mp\varepsilon + \varepsilon^2 \mp\varepsilon^3 +
\varepsilon^4 \mp \varepsilon^5 + \ldots \\
{\displaystyle \frac{1}{(1\pm\varepsilon)^2}} \approx 1 \mp 2\varepsilon +3\varepsilon^2 \mp 4 \varepsilon^3 +
5 \varepsilon^4 \mp 6 \varepsilon^5 + \ldots
\end{array}
\right.
\end{equation}
where, for truncating the expansion up to a certain degree in
$\varepsilon$,  some hypotheses on the smallness of $\varepsilon$
(in comparison with the unity) are required.

In this way, terms of the type $m_1r_1^n \pm m_2r_2^n$ will
appear.

Finally, it is convenient to consider as a basic geometric
configuration, the one shown in Fig.~\ref{gwsl}, where  the plane
of the binary star orbit includes the straight line joining the
source of the ray--light under scrutiny and the observer. That
very configuration is the one which allows the maximum effect.

After each of the following two sub--section, dedicated to the most relevant approximation schemes, we
shall briefly discuss to what extent our statements have to be weakened for a generic geometric configuration.
On the contrary, in Sec. 2.7 hereafter, devoted to an overview of the
 the results,  we will neglect such dependencies, since we are interested in esteeming
 the order of magnitude of light ray deflection. Of course,
these considerations should properly be taken into account for
any specific case.

\subsection{The quasi--static field}

We can now write down the overall deflection of the light beam due to the two masses as the linear superposition
of the light deflection produced by each body. We write such an angle as $\alpha_{\mathrm{ PN}}$,
to distinguish from other sources of deflections that we shall examine in the paper. The angle will be given by:

\begin{equation}
\alpha_{\mathrm{ PN}} = -\frac{4G}{c^2} \left(\frac{m_1}{d_1} + \frac{m_2}{d_2} \right)
\end{equation}
which, using Eq.~(\ref{d1d2}), translates into:

\begin{equation}
\alpha_{\mathrm{PN}} = -\frac{4G}{c^2 d} \left( \frac{m_1}{ 1+\frac{r_1}{d}\cos\omega t} +
                                   \frac{m_2}{ 1-\frac{r_2}{d}\cos\omega t}\right)\,.
\end{equation}
Adopting the first of the expansions in Eq.~(\ref{v27}), the previous relation can be rewritten as:

\begin{eqnarray}
\alpha_{\mathrm{ PN}} & \approx & -\frac{4G}{c^2} \left[(m_1+m_2) \frac{1}{d}
 + \right. \nonumber \\
& & \hspace{2em} + \left(m_1r_1^2 + m_2r_2^2\right)\frac{\cos^2\omega t}{d^3} \\
& & \hspace{2em} \left. - \left(m_1r_1^3 - m_2r_2^3\right) \frac{\cos^3\omega t}{d^4} + \ldots\right] \nonumber
\end{eqnarray}
or in full form:

\begin{eqnarray} \label{v34}
\alpha_{\mathrm {PN}} & = & -\frac{4G}{c^2} \left[ \frac{m_1+m_2}{d} +
\right. \nonumber\\
&& \hspace{2em} +\sum_{n=1}^{\infty}
\left( m_1r_1^{2n}+m_2r_2^{2n}\right) \frac{\cos^{2n}\omega
t}{d^{2n+1}} +  \\
& & \hspace{2em} \left. -\sum_{n=1}^\infty
\left(m_1r_1^{2n+1}-m_2r_2^{2n+1}\right) \frac{\cos^{2n+1}\omega
t}{d^{2n+2}} \right] \;.\nonumber
\end{eqnarray}

We just stress that the first term of the first series expansion
in brackets is not time dependent. It simply gives the deflection
due to the whole mass of the binary, as concentrated in its centre
of mass.
We also note that in case of a perfectly symmetric
binary, only terms depending upon odd powers of $d$ are non zero,
with the $d^{-1}$ term not depending upon the time.

As pointed out by Kaiser \& Jaffe (1997), in some occasions the simple
Schwarzschild metric perturbation (the same effect giving the $\approx
1.75"$ deflection of light at the edge of the Sun) can be of a similar
order of magnitude.
It is clear, however, that the latter statement can be proven only
under some particular conditions.
In fact the larger is the separation of a binary star
the weaker is the GW strength and the stronger is the Schwarzschild metric
perturbation. It has also to be pointed out that while for an oscillating
mass the Schwarzschild perturbation goes down with $d^{-2}$, in the case of a
binary source the perturbation will goes down with a much faster $d^{-3}$
law.

The first time--varying term of Eq.~(\ref{v34}) can be expressed
in astrophysical units as:

\begin{equation}
\alpha_{\mathrm {PN}}^{\prime\prime} \approx 1.75\cdot 10^{-7} \frac{\frac{m_1}{M_\odot}
\left(\frac{\rho_1}{R_\odot} \right)^2 + \frac{M_2}{M_\odot} \left(
\frac{\rho_2}{R_\odot}\right)^2}{\left(d/{\mathrm{ AU}} \right)^3}\,.
\end{equation}
\begin{figure*}
\begin{minipage}{179mm}
\includegraphics[width=179mm]{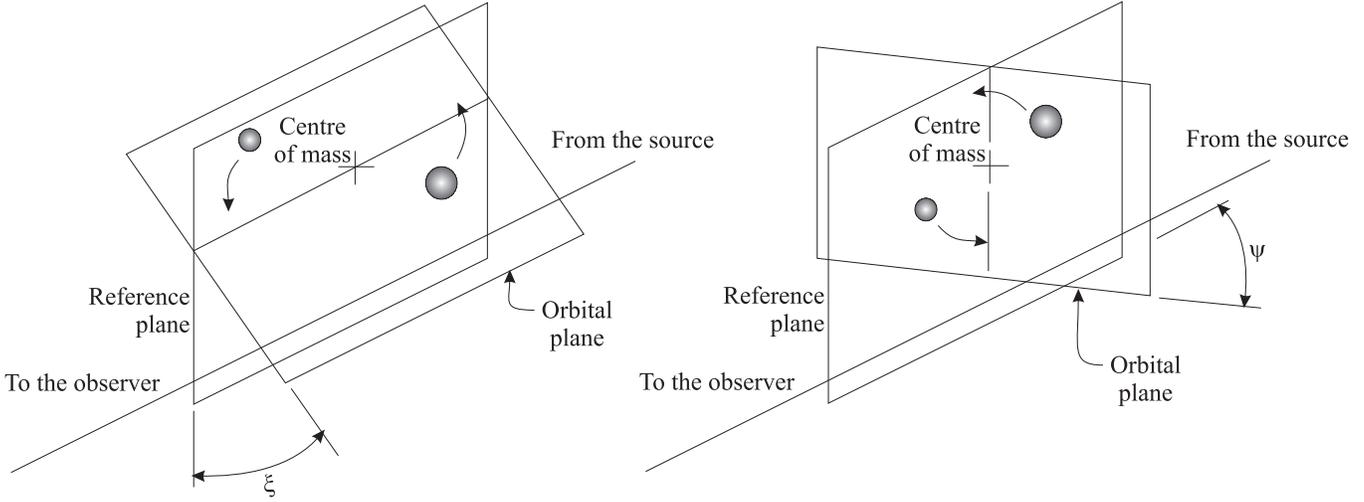}
\caption{ \label{fignew}
Let us define the reference plane  as the
one containing the centre of mass of the binary star and the line
connecting the observer to the source. A generic configuration for
the plane of the orbit may be characterized by the couple of
angles $(\xi,\psi)$. All the basic calculations are carried out
under the simplifying assumption that the two planes coincide.}
\end{minipage}
\end{figure*}

With reference to Fig.~\ref{fignew}, we note that this effect
scales with the cosine of the angle $\xi$. In fact the deflection
above mentioned  disappears when the impact parameter, as seen by
the observer, during motion remains perpendicular to the line
joining the stars of the the binary.

\subsection{A post--Minkowskian treatment}

The deflection angle of an approaching light beam by a single mass is given by the usual $4Gm/c^2$ times $d^{-1}$ in the perturbing mass reference frame.
If the mass is moving with an arbitrary speed $v$, one further term appears, where only the speed component $v_k$ along the line of sight is relevant.
One can can think of it as an additional deflection angle $\alpha_{\mathrm{ PM}}$ given by the linear superposition of the two moving masses in the binary star:

\begin{equation}
\alpha_{\mathrm{ PM}}= \frac{4G}{c^3} \left( \frac{m_1V_{1k}}{d_1} + \frac{m_2V_{2k}}{d_2} \right)\,.
\end{equation}
With the perturbative approach so far described, the latter equation can be rewritten as:

\begin{eqnarray}
\alpha_{\mathrm{ PM}} & = &-\frac{4G \omega \cos \omega t}{dc^3} \left[\frac{m_1r_1}{1+(r_1/d)\cos\omega t} + \nonumber \right. \\
&  &\hspace{7em} -\left. \frac{m_2r_2}{1-(r_2/d)\cos\omega t} \right]
\end{eqnarray}
and expanded as:

\begin{eqnarray}
\alpha_{\mathrm{ PM}} & \approx & \frac{4G\omega}{c^3} \left[\left(m_1r_1^2+m_2r_2^2 \right)\frac{\cos^2\omega t}{d^2} + \right. \nonumber \\
& & \hspace{2em} -\left(m_1r_1^3-m_2r_2^3\right)\frac{\cos^3\omega t}{d^3} + \\
& & \left. \hspace{2em} + \frac{\left(m_1r_1^4 + m_2r_2^4 \right) \cos^4 \omega t}{d^4} + \ldots \right] \,. \nonumber
\end{eqnarray}
The resulting relations can be rewritten in the following compact and exact form:

\begin{eqnarray}
\alpha_{\mathrm{ PM}} & = & \frac{4G\omega}{c^3} \left[ \sum_{n=1}^\infty \left(m_1r_1^{2n} + m_2r_2^{2n} \right)
\frac{\cos^{2n}\omega t}{d^{2n}} + \right. \nonumber \\
& & \left. -\sum_{n=1}^\infty \left(m_1r_1^{2n+1}-m_2r_2^{2n+1} \right) \frac{\cos^{2n+1}\omega t}{d^{2n+1}} \right]\,.
\end{eqnarray}

In contrast to what happens for the Post--Newtonian contribution,
the above effect  depends on the cosine of the angle $\psi$. This
means that, while for the configuration shown in Fig.~\ref{gwsl}
both the PN and PM terms  amount to the full figure worked out
here, for a generic configuration of the orbital plane (see
Fig.~\ref{fignew} for two particular sets of cases, where only one of the two angles is different from zero), these two terms are, in general, attenuated,
but they cannot simultaneously vanish.

\subsection{Post--post--Newtonian relativistic deflection}

Post--Newtonian relativistic deflection by a mass is obtained from the first term expansion of the Schwarzschild metric in the impact
distance $d$. Of course, it is possible to go further and write down the deflection angle up to the $d^{-2}$ term as in Epstein \& Shapiro (1980), see also Ebina et al. (2000). Let us write this additional contribution to the deflection angle, considering, as
before, the linear superposition of the effect due to the two masses in the binary:

\begin{equation}
\alpha_{\mathrm{ PPN}} = -\frac{15\pi G^2}{4c^4} \left[ \left(\frac{m_1}{d_1}\right)^2 +
\left(\frac{m_2}{d_2}\right)^2\right]\;.
\end{equation}
Following the same approach we used in the previous sections, the latter equation can be rewritten as:

\begin{eqnarray}
\alpha_{\mathrm{ PPN}}  =  -\frac{15\pi G^2}{4c^4 d^2} &  \left[
{\displaystyle \frac{m_1^2}{\left( 1+ r_1/d \cos\omega t\right)^2}} + \right. \nonumber \\
  & +\left. {\displaystyle \frac{m_2^2}{\left(1-r_2/d \cos\omega t \right)^2}} \right]
\end{eqnarray}
and, after substituting the expansion given in Eq.~(\ref{v27}), one obtains:

\begin{eqnarray}
\alpha_{\mathrm{ PPN}} & \approx & \frac{15\pi G^2}{4c^4} \left[\left(m_1^2+m_2^2 \right)\frac{1}{d^2} + \right. \nonumber \\
 & & \hspace{2em} -2\left(m_1^2r_1-m_2^2r_2\right) \frac{\cos\omega t}{d^3} + \\
& &  \hspace{2em} \left. + 3 \left(m_1^2r_1^2+m_2r_2^2\right) \frac{\cos^2\omega t}{d^4} + \ldots
\right] \,.\nonumber
\end{eqnarray}
Contrary to what one could expect, it is remarkable that the first non--vanishing,
time--varying term is of the same order as it occurs in the post--Newtonian approach,
at least when a non--symmetric binary is considered.
As before, we also give the complete expression for $\alpha_{\mathrm{PPN}}$:

\begin{eqnarray}
\alpha_{\mathrm{PPN}} & = & \frac{15\pi G^2}{4c^4} \left[ \frac{m_1^2+m_2^2}{d^2} + \right. \nonumber \\
& &\! \! \! - \sum_{n=1}^\infty
2n \left( m_1^2 r_1^{2n-1} - m_2^2r_2^{2n-1}\right) \frac{\cos^{2n-1}\omega t}{d^{2n+1}}+
\nonumber \\
& & \! \! \! \left. + \sum_{n=1}^\infty (2n+1) \left(m_1^2r_1^{2n}+m_2^2r_2^{2n} \right)
\frac{\cos^{2n}\omega t}{d^{2n+2}}\right]\,.
\end{eqnarray}

\subsection{Radiative term cancellation in GR}

The first estimate for the light deflection angle due to GWs is likely to be due to
Fakir (1994a). Under some specific assumptions, he obtains
that, at an impact parameter $d$ equal to one GW wavelength,
a light ray is deflected by an angle $\alpha_{\mathrm{ F}}$ given by:

\begin{equation}
\left.\alpha_{\mathrm{ F}}\right|_{d=\Lambda} = \frac{3}{2} \pi^2 \left. h\right|_{d=\Lambda}.
\label{deltafi}
\end{equation}
This result and its possible extensions to different values of $d$
were examined by many authors (Linet \& Tourrenc, 1976; Durrer, 1994; Fakir, 1995; Kaiser \& Jaffe, 1997),
with results only partially in agreement among themselves.

In particular,
Durrer (1994)
claims $\alpha_{\mathrm{ F}} \propto h$ at any distance $d$.
Under the above assumption, replacing $h$ in Eq.~(\ref{deltafi}) with $h_0$ expressed in Eq.~(\ref{accazero}), one gets:

\begin{equation}
\left. \alpha_{\mathrm{ F}} \right|_{d=\Lambda}=  \frac{12\pi}{\sqrt{5}} \frac{G^{5/3} \omega^{5/3} \mu M^{2/3}}{c^5}
\end{equation}
and, expressing the relationship in MKS units:

\begin{equation}
\left.\alpha_{\mathrm{ F}}  \right|_{d=\Lambda} \approx 7.63 \cdot 10^{-59} \omega^{5/3} \mu M^{2/3}\;,
\end{equation}
where $\alpha_{\mathrm{ F}}$ is given in radians, while in astrophysical units:

\begin{equation}
\left. \alpha_{\mathrm{ F}}^{\prime\prime} \right|_{d=\Lambda}\approx 6.28 \cdot 10^{-10} \frac{\left(\mu/M_\odot \right)
\left(M/M_\odot \right)^{2/3}}{\left( P/{\mathrm{ days}}\right)^{5/3}}\;,
\end{equation}
for deflection angle expressed in arcsec units.
However, Fakir (1994a)
points out a slightly faster decrease of the light deflection with the
distance $d$, being $h\propto d^{-1}$.
Kayser \& Jaffe (1997) confirm Fakir's result for a range of $d$ of the
order of $\Lambda$ but are unable to confirm Durrer's claim.
In more recent times Fakir's results have been strongly criticized.
Bracco (1998) was the first to point out that the dependence from $d^{-1}$ in the
deflection angle is too optimistic,
although he still considered the $d^{-3}$ term he found in its place, as
a radiative one, or, in other words, intimately linked to the GW
nature of the perturbation.

Later, Damour \& Esposito--Farese (1998) and Kopeikin et al. (1999), while pointing out
the same result shown by Bracco, also claimed that the $d^{-3}$ contribution is of
quasi--static nature and has nothing to do with the radiative nature of GW emitted
by the binary. They pointed out a radiative term of the deflection, but only due to
the value of the metric perturbation at the observer's and at the source's location
(called edge--effects), that, of course, prevents to sense GW fields in positions
significantly closer to the very GW source.

While the limits of these findings are going to be briefly
discussed in the next
 sub--section, we want to point out that both Damour \& Esposito--Farese (1998)
 and Kopeikin stress that
 their results are due to a perfect cancellation of terms in $ d^{-1}$ in the GR framework,
 so that any discrepancy in the latter can translate into a renaissance of
 the $\alpha_{\mathrm{ F}}$ term. In particular, scalar GWs are likely to introduce
 radiative deflection angles, which, under particular circumstances,
 could become comparable to the mentioned one, so that one can
 imagine to use a measure of $\alpha$ in order to establish the existence of
 a term of the $\alpha_{\mathrm{ F}}$ type (Faraoni, 1996; Bracco \& Teyssandier,
 1998; Liu \& Overduin, 2000; Will, 2001).

\subsection{Are we neglecting a relevant term?}

Let us now define the average mass $m=(m_1+m_2)/2$ and
the mass asimmetry $\Delta m= |m_2 - m_1|$ of the binary star.
Even if almost all real binaries, like the
one reported in our example (Sec. 5), are not close to be symmetric,
a series expansion in $\Delta m/m$ is still possible,
provided that it is not stopped at the first terms.
Notwithstanding, the first terms can be useful to identify the
different $d^{-n}$ dependencies in the
deviation angles
computed according to the different approximation schemes we examined.
Thus, for the reader's benefit, we have collected the above results
in Tab.~1.

We note that the only non-static term in $d^{-1}$
power is the one claimed by Fakir (1994a). Neglecting such a term,
to find an explicit
 time dependence, it is necessary at least to look at $d^{-2}$ terms.
 It is worth noting that Bracco (1998),
 Damour \& Esposito--Farese (1998) and Kopeikin (1999) not only have pointed out that
 Fakir's result is not
 reliable, but also that the first relevant time dependent term is in
 power of $d^{-3}$, corresponding to our first time dependent PN
 term. They correctly state that such contribution is actually a quasi
 static one and have no radiative nature.
We also point out that Kopeikin's approach not only gives an
independent theoretical confirmation to Damour \&
Esposito--Farese's result, but also extends it to a more general
setting.
 However, our first time dependent
 PM term, clearly is of  non quasi static nature and it is in $d^{-2}$
 power. It appears that such a term is missing both in Damour \& Esposito--Farese (1998) and Kopeikin (1999) works, the first because they explicitely assume that the source internal motions are non--relativistic, so that the time--dependent external gravity field is quadrupolar; the second
because their approximation ``accounts for the static monopole, spin, and time--dependent quadrupole
moments of an isolated system''.
  Moreover, we just
 recall
 that we derived such a contribution through a simple series expansion in $d^{-n}$ terms of
 an equation appearing in  a later work by
 Kopeikin \& Sch\"afer (1999), although the same result is obtainable in the same way
 starting from a
 work by Pyne \& Birkinshaw (1993), recently confirmed with a completely different approach by Frittelli (2003).
For a discussion on the various approximations methods for the investigation of GWs, the interested reader can refer to Thorne (1987) and Zakharov (1973).

If the PM $d^{-2}$  term has really to be interpreted as a
 radiative or even a gravitomagnetic one is a question we do not
 intend to address here.
 We now define a simplified form for the total PM
 ({\em i.e.} including PN but neglecting PPN,  hence developing the result as the first--order term in $h$)
 time depending light deflection
 angle,
 $[\Delta \phi]$, in which one has to take into account that:
 \begin{itemize}
 \item[a):] we consider an expansion both in the GW perturbation $h$
 (see Eq.~(\ref{equa20})) and in the ratio $\Lambda/d$ between the
 characteristic wavelength $\Lambda$  (see Eq.~(\ref{equa25})) and
 the impact parameter $d$;
 \item[b):] just to help the reader in grasping the dependency
 structure of such an expansion, we intentionally omit all the
 trigonometric functions appearing in its time dependent terms;
 \item[c):] the static (not depending on time) terms are
 neglected.
 \end{itemize}
As a result, one obtains:

\begin{table*}
\centerline{
\begin{tabular}{ccccc}
\hline
\hline
    &          &          &          &                        \\
$|\alpha|$  & $d^{-1}$ & $d^{-2}$ & $d^{-3}$ & $d^{-4}$               \\
    &          &          &          &                        \\
\hline
    &          &          &          &                        \\
PN  & Static   & --       & $mr^2$   & $4\Delta m r^3$        \\
    &          &          &          &                        \\
PM  &   --     & ${\omega m r^2}/{c}$ & ${4\omega \Delta m r^3}/{c}$  & ${\omega m r^4}/{c}$ \\
    &          &          &          &                        \\
PPN $\times  \left[15\pi G/\left( 16c^2 \right)\right]^{-1}$ &   --     & Static & $4m\Delta m r$ &  $3m^2r^2$    \\
    &          &          &          &                         \\
F   &  $\approx \left(3\pi^2\omega^2\right)/\left(\sqrt{5}c^2\right) \times mr^2$
               &    ?     &     ?    &      ?                  \\
    &          &          &          &                         \\
\hline
\hline
\end{tabular}
}
\caption{Weights of the various terms with time dependencies accordingly to the various development lines. An overall $4G/c^2$ term is removed. Also note the alternate dependencies upon the system mass $m$ and
the non--symmetric term $\Delta m$. }
\end{table*}

\begin{eqnarray}
[\Delta \phi ] & = &
\frac{\sqrt{5}}{\pi}\,h \left\{ \frac{3 \pi^3}{2
 \sqrt{5}} \, \eta_{\mathrm{ F}} + \right.\nonumber\\
 && +\left(\frac{\Lambda}{d}\right)\left[1+ 2 \frac{\Delta m}{m}\frac{r}{d}+\left(
  \frac{r}{d}\right)^{2}+\ldots\right] +\label{sviluppoY}\\
 &&\left. \!\! - \frac{1}{\pi}
 \left( \frac{\Lambda}{d} \right)^2 \left[1+ 2 \frac{\Delta m}{m}\frac{r}{d}+\left(
  \frac{r}{d}\right)^{2}+\ldots \right] + \ldots
 \right\}\,,   \nonumber
\end{eqnarray}
where $\eta_{\mathrm{ F}}$ is just a coefficient related to  Fakir's claim
that turn out to be zero in the strict GR framework.

The validity of the further expansion in $r/d$, carried out inside each term, resides on the fact that $d\approx \Lambda$ for the cases of interest here (the ones where the various terms in $(\Lambda/d)^\chi$ are comparable for various values of $\chi$) and that $r$ can be obtained through Eq.(\ref{equa25}), recalling that $v=\omega \rho$; in fact, it turns out that the $r/d$ series expansion given here is equivalent to a series expansion in $v/c$.

Such an expression can easily be re-written in the approximation
of $r/d \rightarrow 0$ (small binary star approximation, where the
star dimension $r$ is negligible if compared to the impact parameter
$d$ ) as:

\begin{equation}
 \lim_{\frac{r}{d} \rightarrow 0} [\Delta \phi] = \frac{\sqrt{5}}{\pi}\,h
 \left[ \frac{3 \pi^3}{2
 \sqrt{5}} \, \eta_{\mathrm{ F}} + \frac{\Lambda}{d} - \frac{1}{\pi}
 \left( \frac{\Lambda}{d} \right)^2 + \ldots \right]\,,    \label{finaldeflection}
\end{equation}
which may be clearly interpreted as a series expansion in
$\Lambda/d$. Furthermore, we point out a fact that, in our opinion,
holds true in all the calculations reported here and in the past
(by Fakir 1993, Faraoni 1996, Bracco \& Teyssandier 1998, Bracco
1998, Damour \& Esposito-Farese 1998, Kopeikin et al. 1999).

As one can see from Eq.~(\ref{sviluppoY}), when $\Lambda$ becomes
of the same order than $d$ ({\em i.e.} in the limit $\Lambda/d
\rightarrow 1$), all the terms deriving from the different
approaches (PM, PN ...) to solve Einstein field equations become
comparable in strength. Just some numerical coefficients appear in
the relative ratios, like the $\pi$ factor between the $d^{-2}$
and $d^{-3}$ terms. That means that, in our opinion, further
theoretical developments could lead to terms in $\Lambda/d$ of
order greater than $2$, but still having a relevance in all the
cases where $d$ is of the same order of $\Lambda$. We recall again
that $h \propto d^{-1}$ having as a consequence a dependency upon
$d$ with a power law steeper than a cubic one. Our position is
just that if a certain approximation technique makes a radiative
term disappear up to a certain power, it is necessary to take into
account the first non-vanishing term (if it exists).

That is why it could become particularly important to find an
astrophysical case for probing the situation described above,
that is when higher order terms become relevant.

\section{Perturbation on a microlensing event}

As pointed out in a similar situation by Bracco (1997), some
bending of light of a given angle does not necessarily translates
into effects that are simply proportional to such an angle but
they have to be re--scaled accordingly to the involved geometry.

Let us consider a microlensing event, where
the source $S$, the lens $L$ and the Earth as Observing point, $O$, are laying
approximately on a straight line.
Let us denote by $D_{\mathrm{SL}}$ the distance  between the source and the lens,
by $D_{\mathrm{LO}}$
the one between the lens and the observer, and by $D_{\mathrm{SO}}=D_{\mathrm{SL}} + D_{\mathrm{LO}}$
the distance between the observer and the source.  The impact parameter,  $r$, is
the separation between the lens and the straight line joining the observer and
the source: it is measured on the lens plane orthogonal to the line of sight.
The GW is crossing transversally the line of sight at a distance
$D_{\mathrm{GW}}$ from the Observer. Let us also  assume, hereafter, that the GW wavelength, denoted by $\Lambda$,
is much
greater than both $r$ and $r_E$, the impact parameter and the Einstein
radius of the
microlensing event, respectively. In this way, it becomes negligible
any differential deflection
between different rays focussed by the lens toward the observer.

\subsection{GW source between the Lens and the Observer}

\begin{figure}
\resizebox{\hsize}{!}{\includegraphics{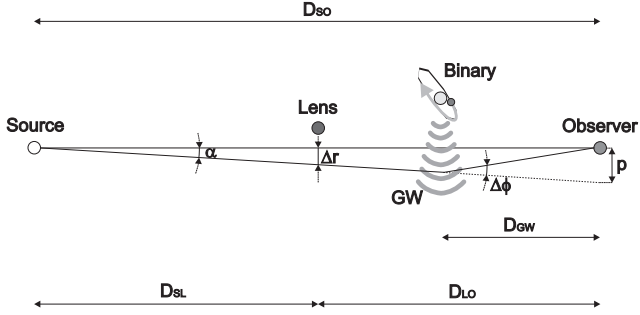}}
\caption{\label{gwlo} A binary star as a source of GWs is between
the lens and the observer during a micorlensing event. GWs will
perturbate the alignment leading to some noticeable signature in
the observed lightcurve. In order to make the figure clearer the
deflection due to the lens is not shown.}
\end{figure}

According to the notations explained in Fig.~\ref{gwlo}, let us
define an auxiliary distance $p$ on the plane defined by the
observer and perpendicular to the microlensing alignment line.
Since the deflection angle is very small, we write

\begin{equation}
p=\Delta \phi\, D_{\mathrm{GW}}
\end{equation}
and, using $\alpha = p/D_{\mathrm{SO}}$ and $\Delta r= \alpha\, D_{\mathrm{SL}}$, one gets

\begin{equation}
\Delta r = \Delta \phi \,\frac{D_{\mathrm{GW}}\,D_{\mathrm{SL}}}{D_{\mathrm{SO}}}\,,
\label{deltar1}
\end{equation}
where $\Delta r$ is the variation of the impact parameter $r$, due to
the GW. Of course such a parameter oscillates, in the case of
a monocromatic GW, between $r-\Delta r$ and $r+\Delta r$. We just
note that when $D_{\mathrm{GW}}=0$, $\Delta r$ vanishes, so that any GW
source close to the observer does not produce any effect. The
maximum effect holds when $D_{\mathrm{GW}}=D_{\mathrm{LO}}$, that is when the GW is
generated in the neighbourhoods of the lens. In such a case,
 the following equation holds true:

\begin{equation}
\Delta r = \Delta \phi \, \frac{D_{\mathrm{LO}} D_{\mathrm{SL}}}{D_{\mathrm{SO}}}\,.
\end{equation}

When $D=D_{\mathrm{LO}}=D_{\mathrm{SL}}$ we obtain $\Delta r =\Delta \phi
\, D/2$, so the maximum optical {\it lever} is equal to one fourth of
the whole distance between the source and the observer. It is worth noting
that, in the case of Galactic measurements, this condition poses a somewhat upper limit
on the maximum lever, given by roughly one half of the  distance
of the Sun from the Galactic center.

\subsection{GW source between the Source and the Lens}

\begin{figure}
\resizebox{\hsize}{!}{\includegraphics{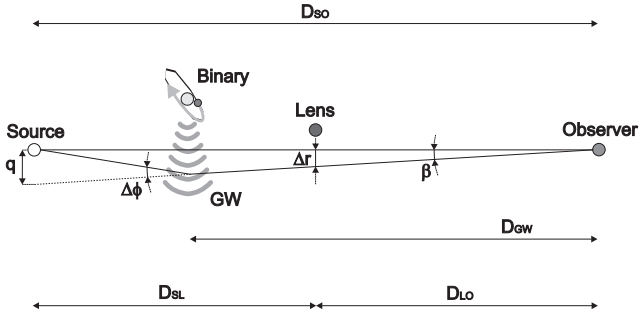}}
\caption{\label{gwsl2} As in Fig.~\ref{gwsl2}, but with the GW's
source between the source and the lens. Again, the deflection due
to the lens is not shown.}
\end{figure}

With reference to Fig.~\ref{gwsl2} it is useful, in this case, to introduce the
displacement $q$, given by

\begin{equation}
q \approx \Delta \phi \left( D_{\mathrm{ SO}} - D_{\mathrm{ GW}}\right)\,;
\end{equation}
as in the previous subsection, let us define the angle $\beta \approx
q/D_{\mathrm{ SO}}$ and the variation $\Delta r \approx \beta \, D_{\mathrm{ LO}}$, thus obtaining

\begin{equation} \label{deltx}
\Delta r = \Delta \phi \,  \frac{D_{\mathrm{ SO}}- D_{\mathrm{ GW}}}{D_{\mathrm{ SO}}} \, D_{\mathrm{ LO}} \,.
\end{equation}
It is easy to see that the above relation is close to Eq.~(\ref{deltar1}).
The behaviour of the optical lever is similar, provided that one replaces
the distance of
the GW from the Observer with the same distance, but measured from the source.
It is also evident that Eq.~(\ref{deltx}) and Eq.~(\ref{deltar1})  give the same value
when the GW is located close to the lens. The same considerations described in Section
3.1 hold true here.

With reference to Fig.~\ref{lgw}, one can introduce a lever length $l_{\mathrm{ GW}}$,
such that $\Delta r = \Delta \phi \, l_{\mathrm{ GW}}$.
The behaviour of $l_{\mathrm{ GW}}$, measured on the microlensing straight line connecting
the source and the observer is the following one: starting from the zero value at the source,
it linearly grows since it reaches its maximum at the lens position, then it linearly
decreases to zero,
approaching the observer.
Provided that a certain sensitivity to $\Delta r$ is accomplished,
a bi--conic volume of the Galaxy is probed in searching for GWs larger than a
given threshold (see the upper right insert in Fig.~\ref{lgw}).

\section{Photometric effects on microlensing}

Following Paczinsky (1986) we use the Einstein radius $r_E$ defined as

\begin{equation} \label{v35}
r_{\mathrm{ E}} = \sqrt{\frac{4GM_{\mathrm{ L}}}{c^2}\,\frac{D_{\mathrm{ LO}} D_{\mathrm{ LS}}}{D_{\mathrm{
S}}}}\,,
\end{equation}
where $M_L$ is the mass of the lensing object. Denoting by $u=r/r_E$,
a dimensionless impact parameter,  the amplification factor $A(u)$
is given by:

\begin{equation}
A(u) = \frac{u^2+2}{u\sqrt{u^2+4}}\,.
\end{equation}
The GW within the line of sight of a microlensing event introduces a tiny
perturbation in $u$ and a corresponding modulation of the amplification,
which can be esteemed to be:

\begin{equation}
\frac{\partial A}{\partial u} = -\frac{8}{u^2(u^2+4)\sqrt{u^2+4}}\,.
\end{equation}
We point out that in the cases of interest here, $u \ll 1$ holds true,  hence the
previous relations  simplify into $A\approx u^{-1}$ and
$\partial A/\partial u \approx -A^2$.

\begin{figure}
\resizebox{\hsize}{!}{\includegraphics{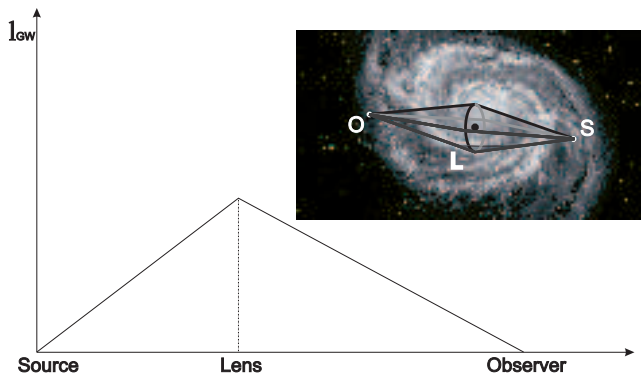}} \caption{The
ratio between the effect on the impact parameter, $\Delta r$, and
the deflection angle generated by the GW perturbation, $\Delta
\phi$, is here defined as a {\it lever} length $l_{\mathrm{ GW}}$
which is maximum near the lens and drops linearly to zero both
toward the source and the observer. For a hypothetical event,
where the lensing object is located in the bulge of our Galaxy
(see upper--right inset), the searching for a GW signature as
described in the text is equivalent to probing for GW sources
within a biconical volume. \label{lgw}}

\end{figure}

Due to the perturbation, $\Delta r$ the dimensionless impact parameter is perturbed
by an amount $\Delta u = \Delta r /r_E$ and the magnification $A$ will exhibit, in the approximation $A \gg 1$:

\begin{equation}
\Delta A \approx -A^2 \Delta u\,.
\label{deltaa}
\end{equation}
Such a perturbation leads to a hopefully measurable brightness
variation $\Delta I$ of the observed flux $I$ (see also
Fig.~\ref{lgw}). Let us define $\Delta m$ as the maximum
photometric magnitude difference between several measurements
affected by an intensity $I \pm \Delta I$, {\em i.e.}:

\begin{equation}
\Delta m = \frac{5}{2} \log \left( 1 + 2 \frac{\Delta I}{I}
\right)\,,
\end{equation}
so that one can approximately write:

\begin{equation}
\Delta m \approx \frac{5}{\ln 10} \frac{\Delta A}{A}\,.
\end{equation}
By using Eq.~(\ref{deltaa}) and the concept of effective length $l_{GW}$
defined in the previous section, one can write

\begin{equation}
\Delta m \approx \frac{5}{\ln 10} \frac{\Delta \phi A\, l_{GW}}{r_E}\,.
\label{deltam}
\end{equation}
Taking into account Eq.~(\ref{accazero}), Eq.~(\ref{deltafi}) and Eq.~(\ref{deltam})
it is possible to estimate the maximum projected distance $d_{\mathrm{ max}}$, around which  a binary
star  produces a perturbation, leading to a given photometric amplitude $\Delta m$:

\begin{equation}
d_{\mathrm{ max}} = \frac{12 \sqrt{5} \pi^2}{\ln 10} \cdot
\frac{G^{5/3}}{c^4}\cdot  \frac{A l_{GW}}{r_E} \cdot
\frac{\omega^{2/3}\mu M^{2/3}}{\Delta m}\,,
\label{dmax}
\end{equation}
where we have grouped on the right side of Eq.~(\ref{dmax}) the
numerical coefficients, the physical constants, the microlensing parameters
and the binary star's one into four different
fractions.
The linear dependence upon $l_{GW}$ we have found deserves, however, some specific
comments. We just note, in fact, that Eq.~(\ref{v35}) can be rewritten as:

\begin{equation}
r_E = \sqrt{r_S l_\mu}\,,
\end{equation}
where $r_S$ is the Schwarzschild radius of the lens and $l_\mu$ is the
effective length of the microlensing setup, defined similarly to
$l_{GW}$. While $l_{GW}$ and $l_{\mu}$ are {\it a priori}
fully uncorrelated, it is clear that by selection effect, the largest $\Delta
m$ can be obtained when $l_{GW} \approx l_\mu$. Under this condition,
the dependence in Eq.~(\ref{dmax}) from such a
characteristic length becomes weaker, {\em i.e.} a square root.
Also the dependence upon the
mass of the lensing object becomes an inverse square root.

\section{An example with a W--UMa binary}

Following Mironovski (1966) we take a typical W--UMa binary as the average
of the ones listed by Kopal (1959), obtaining  $m_1= 1.46M_\odot$,
$m_2= 0.78 M_\odot$ and $P=0.3^d$, leading to $\omega \approx 2.42 \cdot
10^{-4}s^{-1}$. This kind of double star has a typical size of $\approx 10^{-2}$AU and a radial velocity $v$ such that $v/c\approx 10^{-4}$.
A GW wavelength $\Lambda \approx 26$AU is obtained. We just
note that such a figure is at least an order of magnitude larger than a
typical $r_E$ for microlensing in the Galaxy, so that the approximations
used in the calculation reported in this work appear reasonable.

At a distance $d=\Lambda$, an $h_0\approx 1.33 \cdot 10^{-15}$ is obtained. Assuming
that this W--UMa is in the bulge of our Galaxy, the same perturbation on the
Earth will be lowered to a mere $h \approx 1.68 \cdot 10^{-23}$.

According to Eq.~(\ref{finaldeflection}),  one estimates a light deflection of
$\Delta \phi \approx 9.5 \cdot 10^{-16}$, equivalent to $\Delta
\phi \approx 0.20$ nano--arcseconds (mainly from the PM term).
 Here and in the following
considerations, we assume that what we called Fakir's contribution vanishes:
$\eta_{\mathrm{ F}} =0$; nonetheless, we note that our results could increase by
a factor $\approx 20$ in case  that $\eta_{\mathrm{ F}}=1$.

Let us suppose that a microlensing event occurs and the corresponding
 magnification is relatively high ($A=100$); let us also suppose that
the lens determines a value $r_E=0.1$AU and
the average W--UMa  under investigation is located near the Galaxy center, at
roughly 10kpc from the Sun. With the further assumption that the
source is situated  10kpc far away from the lensing mass, a $l_{\mathrm{
GW}}=5$kpc is obtained. Combining these figures in Eq.~(\ref{deltam})
one can estimate
$\Delta m \approx 2.2 \cdot 10^{-3}$.
Nevertheless, one should note
that a precision of the order of $1/100$ of
magnitude should be enough to detect such a GW event, provided that
one takes the average of several GW--induced oscillations around the time
corresponding to the
magnification peak.

The assumption we made about $r_E$ is roughly one order of
magnitude smaller than a typical one. While we note that this is
just a factor three times smaller than the typical mass for microlensing, we
point out that similar results can be obtained with a smaller
impact factor $d$. In particular, the PN term, growing with
$d^{-3}$, can make a similar perturbation on the microlensing
event when, with a more typical $r_E =1$AU, the impact factor
drops to a $d\approx 5.8$AU, while the PM term, much larger by a
factor $\pi$ at the one GW wavelength distance, behaving  just as
a $d^{-2}$ power,  becomes of the same magnitude at $d\approx
5.1$AU.

\begin{figure}
\resizebox{\hsize}{!}{\includegraphics{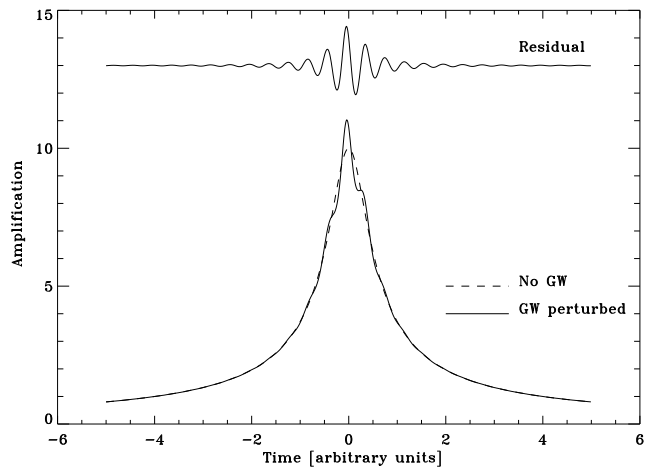}} \caption{An
example of a (strong) signature of GWs generated by a binary star
located close to the line of sight of the micro-lensing event. The
light from the binary source (that would be likely to appear as an
eclipsing binary) is not included in this plot. \label{fig5}}
\end{figure}

All the calculations reported in this paper are carried out under
the assumption that the rays focussed by the microlensing are
approximately subject to the same gravitational deflection. Even
if this condition fits  very well the astrophysical cases
described here, of course it is no longer true when the binary
itself is responsible for microlensing. It is easy to realize that
in this latter case, the effect should be much smaller, because
the focussing deflection would slightly change and the relative
variation in the amplification $A$ should be of the same order of
magnitude as the ratio between the deflection due to the
time--dependent terms and the deflection caused by the static
contributions. Nevertheless, since such a possibility was beyond
the scope of our paper, we have not investigated it in  more
detail, so that we cannot exclude that further interesting results
could be obtained along this way.

\section{Detection probability}
From an inspection of the MACHO project microlensing alert Web
page\footnote{{\tt http://darkstar.astro.washington.edu}} we found
an average detection rate of events with $A>8$ of the order of
$\approx 5$ events per year, with an average light amplification
$\bar A \approx 20$; we dropped from our estimate caustic crossing
events, whose treatment is beyond the scope of this paper. In the
MACHO program $N_* \approx 4.3 \cdot 10^5$ stars are
photometrically observed (Alcock et al., 1995) hence we can define
an Event Detection Rate (hereafter denoted by EDR) $\eta$,
expressed in detected events per star, per year, for such a type
of {\it high--magnification} events, as

\begin{equation}
\eta = \frac{N(A>8)}{N_*} \approx 1.2 \cdot 10^{-5}\,.
\end{equation}

Because $A\approx u^{-1}$ one can easily use the estimation of
$\eta$ to evaluate the EDR corresponding to the case in which
another star (the lens) falls within the line of sight of a given
star in the bulge (the source) at a certain distance
$d_{\mathrm{lensing}}$, given by

\begin{equation}
d_{\mathrm{lensing}} \approx \frac{r_E}{\bar A} \approx 0.05 {\mathrm{AU}}\,.
\end{equation}

\begin{figure}
\resizebox{\hsize}{!}{\includegraphics{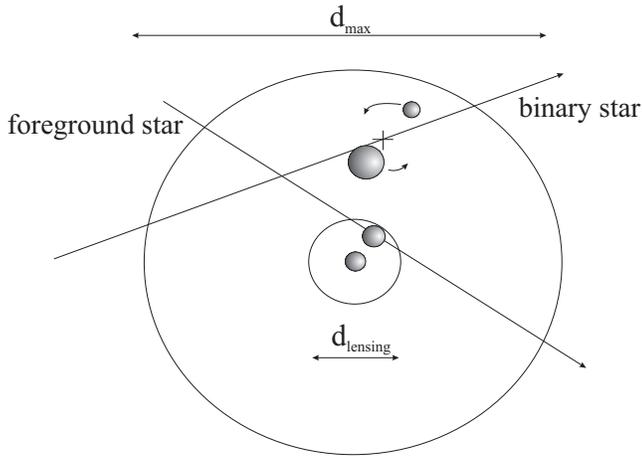}} \caption{In a
reference frame with the origin fixed with the mass causing
microlensing, appearing in the centre of the figure, the
probability that simultaneously a suitable binary is within a
circle of diameter $d_{\rm max}$ and a foreground star is within
another of diameter $d_{\rm lensing}$ is computed in the text.
Moreover,the case corresponding to independent objects is first
evaluated, and then the one in which the mass responsible for the
microlensing is a third companion of the binary is considered.
\label{prob}}
\end{figure}

In fact, such an EDR for a diameter of the order of
$d_{\mathrm{max}}$ will be obtained by simply scaling $\eta$ by
the ratio of the cross-sections defined by the two distances under
study (see Fig.~\ref{prob}), leading to:

\begin{equation}
{\mathcal R} = \eta \left(
\frac{d_{\mathrm{max}}}{d_{\mathrm{lensing}}}\right)^2\,.
\end{equation}
The EDR to have both a microlensing event (like the ones
considered here) and, simultaneously, a third object within a much
larger distance $d_{\mathrm{max}}$ can be written as the product
of the two terms:

\begin{equation}
{\mathcal R} \approx \gamma \eta^2 \left(
\frac{d_{\mathrm{max}}}{d_{\mathrm{lensing}}}\right)^2\,,
\end{equation}
where $\gamma$ is a coefficient that takes into account that the
two events should simultaneously occur.
Indeed, a given star spends a fraction $\gamma \eta$ of the time
undergoing high-magnification events, where $\gamma$ is the
average duration of a high-magnification event, expressed in
years.
A typical value for $\gamma$ is of the order of $10^{-2}$.

Let us estimate $d_{\mathrm{max}}$ for a typical Galactic event, where the lens
is located in the bulge: in this case we assume $l_{GW} \approx 5kpc$
and $r_E \approx 1$AU.
In order to achieve the highest possible photometric accuracy, it
is to be recalled that one can average a few GW periods in the
time span covered during the high--magnification event. Assuming a
final error of the order of $\Delta m \approx 10^{-3}$ mags
(Frandsen, 1993; Gilliland \& Brown, 1992) one can obtain, using
Eq.~(\ref{deltam}) and Eq.~(67), together with the average W--UMa as
from
the previous section, and the above mentioned estimates, a
figure for $d_{\mathrm{max}} \approx 6$AU. We just note this is
significantly smaller than $\Lambda$ so that, in such a
regime, some still unveiled terms
in higher power on $\Lambda/d$ can even dominate.

Finally, one have to further select only the cases where the
second star is of  W--UMa type. The population density of W--UMa
is almost constant all the way to the Galactic bulge, amounting to
roughly $\rho_{\mathrm{W-UMa}} \approx 1/280$ (Rucinski, 1994,
1997). Then, one can combine all these EDR parameters into a
single relationship, giving the average GW-detection time interval
$\tau \approx {\mathcal R}^{-1}$ between two successive events:

\begin{equation}
\tau = \frac{d_{\mathrm{lensing}}^2}{\gamma \eta^2
d_{\mathrm{max}}^2 \rho_{\mathrm{W-UMa}} N_{*}} \approx 4.3 \times
10^{4} \, {\mathrm{ yrs}}\,.
\end{equation}

This figure could make any reasonable search for such event, truly
hopeless. It has been obtained assuming that the W--UMa and the
lensing star position are fully uncorrelated (which translates
into the square power dependency upon $\eta$ and to the appearance
of $\gamma$). However, one should also consider the possibility of
having a triple star system composed by a W--UMa and a third
companion at a much larger distance, responsible for the
microlensing. Because $d_{\mathrm{max}}$ is so small, we found
that the triple star case, in spite of its unlikeness, has an even
higher EDR than the simpler case previously discussed.

In the triple stars case, in fact, the EDR of  a microlensing
event on the companion of a W--UMa, with high magnification rate,
is simply given by:

\begin{equation}
{\mathcal R}' \approx  \eta \rho_{\mathrm{W-UMa}} \eta_3\,,
\end{equation}
where $\eta_3$ is the fraction of W--UMa exhibiting a third
companion at a projected distance of the order of magnitude similar to $d_{\rm max}$. Such a number turns out to be of the order of $\eta_3
\approx 0.2$ (Herczeg, 1988; Tokovinin, 1997), leading to a typical
time interval between two different GW potential detections (or at
least of the gravitationally induced effects discussed in the
text) of the order of:

\begin{equation}
\tau' = \frac{1}{\eta \rho_{\mathrm{W-UMa}} \eta_3 N_*} \approx
273 \, {\mathrm{ yrs}}\,,
\end{equation}
that is  a factor roughly two hundred times smaller than in the case of
{\it free} floating W--UMa and microlensing star. We note that in
the triple star case the condition $L_{\mathrm{GW}} = l_\mu$ is
automatically fulfilled.

Care should be taken into account in considering such figures.
Most of the parameters involved are rough estimates and several
uncertainties of the order of at least a factor two can be
considered. This leads, in our opinion, to a significant
diminishing in the order of magnitude for $\tau'$.

One should also consider that such a time will be lowered by
future improved photometric capabilities, and that it has been devised
for an average W--UMa: in principle one cannot exclude that, by chance,
some stronger event may occur. Moreover, we point out that  the
simultaneity condition discussed in the case of
incorrelation between the W--UMa and the lensing star, is well
verified for the more optimistic assumptions. Notwithstanding, it can
introduce some further augmentation of $\tau$ for the more
pessimistic calculations. The relatively low time--scale, however,
also suggests that there is a small chance that a gravitational
effect signature (to be ascribed to a genuine GW effect or mainly
to a Post--Minkowskian contribution is here beyond our purpose) could
even be hidden in some of the microlensing photometric runs
already collected from the ground.

\section{Conclusions}

In the writers' opinion, a critical review of the light deflection contributions due to the gravitational field of a binary star shows that terms directly linked to a GW (or radiative ones) cannot be excluded by current approximations, available in the literature. When the impact parameter is of the same order of magnitude of the GW's wavelength, such terms could be as important as the leading, non radiative ones. Moreover, a non quasi--static term depending upon the inverse square of the impact parameter is found. Such deflection angles are so small that one would be tempted to leave them in the academic rather than the experimental realm. However, we have shown that such deflection, when occourring close to the line of sight of a
micro--lensing event, can give rise to detectable effects.
The probability of such an alignment is, however, extremely small, except for the case of microlensing by a third wide companion of a close binary star. In that case, from the probability to detect such an event, one can reasonably expect that those effects could be observable in the near future, or even that one or a few of them are hidden in the existing literature data.

We avoided to speculate, in the previous sections, on some extreme cases
where the gravitational effects could be very large. One could, in principle, conceive
a geometry where a small lens (for instance with a mass of the order of Jupiter)
produces a highly amplified  ($A\approx 100$) microlensing event for a duration of several
days, with a massive binary like $\mu$--Sco, close to the line of sight. In this
case, the binary star could form with the microlensing event an angle
sufficient to photometrically study the binary star well separated from the microlensing
event, for a reasonable number of binary periods.
Even if such an event appears to be very unlikely, it
could determine many insights on gravitational and GW studies, including the possibility to
probe gravitation theories that do not lead to a perfect cancellation of
the term claimed by Fakir.

On the other hand, we believe remarkable  that {\it ordinary}
events, where a gravitational or GW signature can be clearly
identified, have a significant chance to pop--up in the current
surveys. Post--microlensing studies should permit to eventually
confirm the GW nature of the waviness detected during
microlensing. Such approaches include spectroscopic and photometric studies
(to ensure in detail the nature of the binary star responsible of
the GWs) and, in future, astrometric studies carried out with some
high angular resolution tool (speckle or adaptive optics) in order
to establish the exact geometry of the three objects involved
(source, lens and binary stars) at the moment of the microlensing
peak magnification.

A positive detection, such as the one described here,
can be seen as a sort of astrometric detection of a GW effect using a Galaxy--sized
telescope whose objective is made by the gravitational lens responsible of the
microlensing (Labeyrie, 1994).

\section*{Acknowledgements}
The authors thank an anonymous Referee for the careful reading of the manuscript
and for many interesting, deep and valuable suggestions and comments. We also
thanks Francesco Sorge for comments and for having brought to his attention the book by Zakharov.

\end{document}